\documentclass[runningheads]{llncs}

 \usepackage{eccv}

\usepackage{eccvabbrv}

\usepackage{graphicx}
\usepackage{booktabs}

\usepackage{amsmath}
\usepackage{amssymb}
\usepackage{cite} 

\usepackage{multirow}
\usepackage{adjustbox}

\usepackage{pifont}

\usepackage[accsupp]{axessibility}  

\usepackage{hyperref}

\usepackage{orcidlink}

\begin{document}

\title{Edit2Restore: Few-Shot Image Restoration via Parameter-Efficient Adaptation of Pre-trained Editing Models} 

\titlerunning{Edit2Restore}

\author{
Mustafa~Ak{\i}n~Y{\i}lmaz\inst{1}\orcidlink{0000-0002-0795-8970}
\thanks{Corresponding author. Code and trained LoRA adapters are available at \url{https://github.com/makinyilmaz/Edit2Restore}.}
\and
Ahmet~Bilican\inst{2}\orcidlink{0009-0000-6406-8168}
\and
Burak~Can~Biner\inst{1}\orcidlink{0009-0009-8580-8220}
\and
Ahmet~Murat~Tekalp\inst{3}\orcidlink{0000-0003-1465-8121}
}
\authorrunning{M.~A.~Y{\i}lmaz et al.}

\institute{
Codeway AI, Istanbul, Turkey
\and
ETH Zurich, Zurich, Switzerland
\and
Ko\c{c} University, Istanbul, Turkey
}

\maketitle

\begin{figure}[t]
  \centering
\includegraphics[width=\textwidth,height=0.25\textheight,]{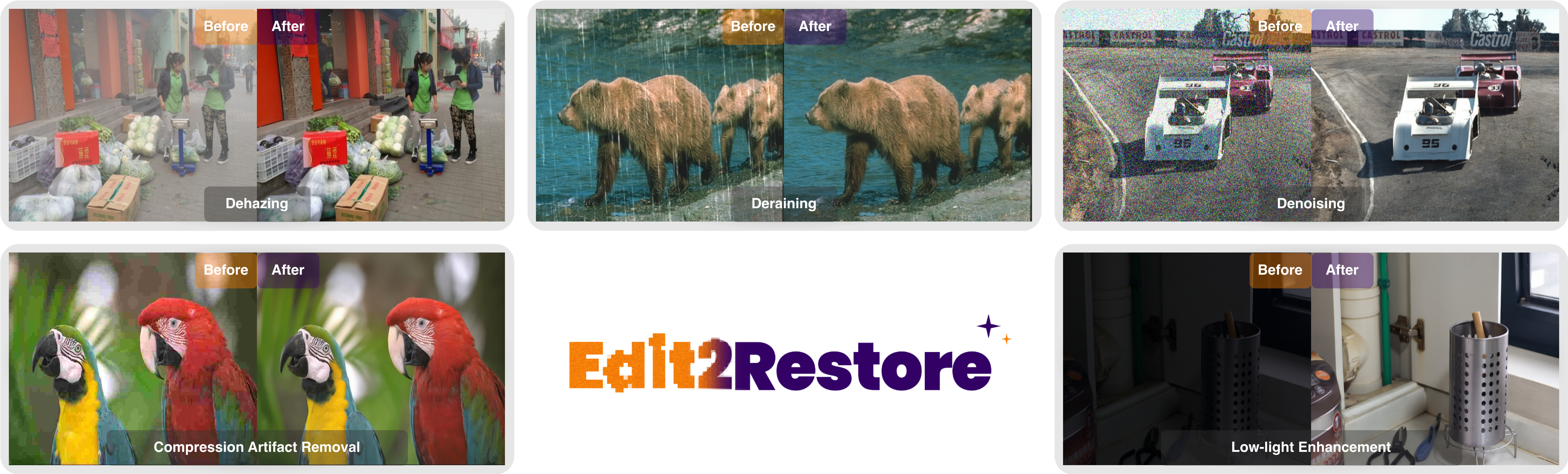}%
  \caption{Edit2Restore effectively restores diverse degradations. Across a single unified model, our method handles dehazing, deraining, denoising, JPEG compression artifact removal, and low-light enhancement.}
  \label{fig:entry}
\end{figure}

\begin{abstract}
  Image restoration has traditionally required training specialized models on thousands of paired examples per degradation type. Large pre-trained text-conditioned image editing models encode rich priors about image structure, quality, and degradation, yet we find that this knowledge does not, on its own, make them restorers: state-of-the-art editing models largely fail at restoration in the zero-shot regime. We show that what these priors lack is not capability but direction, and that a small amount of parameter-efficient adaptation supplies it. Fine-tuning LoRA adapters on FLUX.1 Kontext, a 12B-parameter flow matching model for image-to-image translation, with only 32--128 paired images per task and guided by simple text prompts, we turn a mediocre zero-shot editor into a competitive restorer. A single unified adapter, conditioned on task-specific prompts, handles five diverse degradations. Despite using three to four orders of magnitude less data, our few-shot model surpasses a recent restoration baseline trained on over a million curated pairs on the majority of perceptual and distribution-level metrics, on which we evaluate in keeping with our focus on perceptual rather than pixel-fidelity quality. Through comprehensive studies, we analyze the impact of training-set size, the trade-off between task-specific and unified multi-task adapters, the effect of text encoder adaptation, and zero-shot baseline performance, establishing pre-trained editing models as a compelling, data-efficient foundation for few-shot, prompt-guided image restoration.
\end{abstract}

\section{Introduction}
\label{sec:intro}

Image restoration, recovering high-quality images from degraded observations, has traditionally relied on training specialized models for each degradation type, requiring thousands of paired training examples per task. Recent all-in-one restoration methods~\cite{airnet2022,promptir2023,instructir2024} attempt to handle multiple degradations within a unified framework, yet still demand extensive paired datasets and specialized architectures trained from scratch. For example, InstructIR~\cite{instructir2024} requires training a restoration network with over 10,000 synthetically generated text prompts and thousands of image pairs per degradation to achieve state-of-the-art results. More recently, a complementary direction has begun to repurpose large-scale text-conditioned image editing models for restoration. RealRestorer~\cite{realrestorer}, for instance, fine-tunes an open-source editing model on roughly two million synthetic and web-sourced degraded--clean pairs spanning nine degradation types. While this confirms that editing priors transfer to restoration, it inherits the very data appetite we seek to eliminate, leaving open the question of how little supervision such adaptation actually requires.

In this work, we demonstrate that FLUX.1 Kontext~\cite{fluxkontext2025}, a prominent text-conditioned image editing model based on rectified flow matching, can be efficiently adapted for multiple restoration tasks using remarkably few examples. FLUX.1 Kontext is a 12 billion parameter model designed for image-to-image translation, capable of performing diverse editing operations guided by natural language instructions. Our key insight is that the visual understanding already encoded in such models including knowledge about image quality, degradation patterns, and natural statistics can be efficiently transferred to restoration tasks through parameter-efficient fine-tuning (PEFT), without requiring the model to be trained from scratch.

We fine-tune lightweight LoRA~\cite{lora2022} adapters on FLUX.1 Kontext, guided by simple text prompts that specify the restoration operation. The model performs image-to-image translation: given a degraded input image and a restoration instruction, it generates the corresponding clean output. Unlike InstructIR, which trains a specialized restoration network from scratch, we leverage the pre-trained visual understanding already embedded in large-scale image editing models. A single unified LoRA adapter, conditioned on task-specific text prompts, effectively handles five diverse degradations: denoising, deraining, dehazing, JPEG compression artifact removal, and low-light enhancement, all learned from minimal training data. Crucially, we find that even strong editing models such as Qwen-Image-Edit~\cite{wu2025qwenimagetechnicalreport}, Step1X-Edit~\cite{step1xedit}, FLUX.2~\cite{flux2}, and LongCat-Edit~\cite{longcat} largely fail at these tasks in the zero-shot regime, underscoring that task-specific adaptation is what unlocks restoration.

Unlike traditional image restoration methods that are optimized for pixel-level reconstruction metrics (PSNR/SSIM) using $\ell_1$ or $\ell_2$ losses, our approach optimizes for perceptual image quality, consistent with flow matching-based editing models. Accordingly, we assess restoration with perceptual and distribution-level metrics such as LPIPS, FID, and CMMD rather than pixel-fidelity scores.

Overall, we can summarize our contributions as follows: 
\begin{itemize}
    \item We show that pre-trained text-conditioned image editing models can be adapted for image restoration from only a handful of paired examples, achieving orders-of-magnitude better data efficiency than both from-scratch restoration networks and recent editing-model-based approaches that rely on large-scale curated data.

    \item We introduce a parameter-efficient fine-tuning framework in which a single LoRA adapter handles five restoration tasks (denoising, deraining, dehazing, JPEG compression artifact removal, and low-light enhancement) through natural language instructions, eliminating the need for task-specific architectures.

   \item We conduct extensive ablations examining: the relationship between training set size and restoration quality from 32 to 128 images per task; trade-offs between task-specific and unified multi-task LoRA training; and the impact of LoRA-adapting the CLIP text encoder versus keeping it frozen, while T5 remains frozen throughout.

    \item We provide a comparison against a suite of state-of-the-art editing models in the zero-shot setting and against a strong large-data restoration baseline, demonstrating that our few-shot adapted model surpasses them while prioritizing perceptual quality and data efficiency.
\end{itemize}

\section{Related Work}
\label{sec:related}

\subsection{Task-Specific Image Restoration Models}
Image restoration aims to recover high-quality images from degraded observations. Early research developed specialized models for specific degradation types, including denoising~\cite{DnCNN}, deblurring~\cite{GoPro, DeblurGAN}, low-light enhancement~\cite{lol, RetinexFormer}, dehazing~\cite{FFA-Net}, deraining~\cite{MSFPN}, desnowing~\cite{DesnowNet, SnowFormer}, JPEG compression artifact removal~\cite{ARCNN}, and raindrop removal~\cite{Attentive-GAN}. CNN-based methods like NAFNet~\cite{NafNet}, MPRNet~\cite{MPRNet}, and HINet~\cite{HINet}, followed by transformer-based approaches such as Restormer~\cite{restormer2022}, SwinIR~\cite{swinir2021}, and Uformer~\cite{Uformer} advanced the field through better architectural designs. However, these methods typically optimize pixel-level objectives using $\ell_1$ or $\ell_2$ losses, require thousands of paired examples per task, and primarily target pixel fidelity rather than perceptual quality.

\subsection{Multi-Task Image Restoration Models}

Recent work has explored models capable of handling multiple restoration tasks. Among these, AirNet~\cite{airnet2022} employed contrastive learning to distinguish degradation types. Prompt-based methods~\cite{promptir2023, instrcut-ipt, prompt_in_prompt, ProRes, instructir2024} introduced learnable parameters or natural language instructions to encode degradation-specific information, guiding restoration backbones like NAFNet~\cite{NafNet} and Restormer~\cite{restormer2022}. For instance, InstructIR~\cite{instructir2024} uses human-written text prompts to achieve strong results across multiple tasks. However, these methods train restoration networks from scratch, requiring thousands of paired images per task and substantial computational costs. Only recent work like LoRA-IR~\cite{lorair} has explored parameter-efficient adaptation through low-rank techniques.

With the success of diffusion models in generation, several works leverage diffusion priors for restoration~\cite{DA-CLIP, DiffBIR, autodir, mperceiver, StableSR, DiffRestorer}. While effective, most diffusion-based approaches are computationally expensive and require large data. Most recently, RealRestorer~\cite{realrestorer} fine-tunes a large-scale image editing model (Step1X-Edit) for real-world restoration, constructing a degradation-synthesis pipeline of roughly two million synthetic and web-sourced pairs across nine degradation types. This shares our core motivation of exploiting editing priors, yet operates at the opposite end of the data spectrum: whereas RealRestorer depends on million-scale curated supervision, we demonstrate that as few as 32--128 pairs per task suffice for comparable or better perceptual quality.

\subsection{Text-Conditioned Image Editing Models}

Recent text-conditioned image editing models excel at modifying images according to natural language instructions while preserving unrelated content. Early work like InstructPix2Pix~\cite{brooks2022instructpix2pix} demonstrated instruction-following editing, while recent models achieve state-of-the-art quality: FLUX.1 Kontext~\cite{fluxkontext2025} employs a dual-stream architecture processing image and text separately, and Qwen-Image-Edit~\cite{wu2025qwenimagetechnicalreport} offers a competitive alternative. A rapidly growing family of open instruction-based editors has since emerged: Step1X-Edit~\cite{step1xedit} couples a multimodal LLM with a diffusion decoder, the FLUX.2 family~\cite{flux2} provides compact rectified-flow transformers that unify generation and editing, and LongCat-Edit~\cite{longcat} is a lightweight bilingual editor. Unified models~\cite{chen2025janus, ma2024janusflow, xiao2024omnigen} further handle both understanding and generation within single architectures. While these models form natural zero-shot baselines for restoration, their strong editing priors do not directly translate into reliable restoration capability without task-specific adaptation. Recent perspectives trace the evolution of image priors from handcrafted models to generative foundation models and their use in inverse problems~\cite{tekalp2026imagemodeling}. We extend this direction by arguing that image editing models provide a more natural foundation for restoration.

Our key insight is that image-to-image translation aligns naturally with image-to-image restoration: transforming degraded inputs to clean outputs via text instructions is fundamentally an editing operation. Rather than training from scratch, we demonstrate that parameter-efficient fine-tuning of these pre-trained priors requires only 32--128 examples per task, dramatically reducing data requirements while achieving competitive quality.

\subsection{Parameter-Efficient Fine-Tuning}

Parameter-efficient fine-tuning (PEFT) methods~\cite{zhou2022coop, lora2022, waveft} adapt large pre-trained models to new tasks while updating only a small fraction of their parameters, avoiding the cost and storage overhead of full fine-tuning. Among these, LoRA (Low-Rank Adaptation)~\cite{lora2022} injects trainable low-rank matrices into the weight layers of a frozen backbone, achieving strong performance across language and vision tasks at a fraction of the trainable-parameter budget. Its low-rank updates are particularly attractive for large generative models, where full fine-tuning is prohibitively expensive and risks catastrophic forgetting of the pre-trained priors. LoRA has consequently become a standard tool for adapting large diffusion and flow-based image models, and recent restoration work such as LoRA-IR~\cite{lorair} has begun to exploit low-rank adaptation for efficiency. However, prior efforts still train on large paired datasets and do not exploit the editing priors of large-scale instruction-following models.

In this work, we show that LoRA adaptation of FLUX.1 Kontext, combined with task-specific text conditioning and only a handful of paired images, is sufficient for effective multi-task restoration. Because the backbone remains frozen and only the low-rank adapters are trained, a single set of lightweight weights can serve all five degradation types, keeping the storage and deployment footprint minimal. Building on this, our experiments examine the key factors governing such adaptation: the amount of paired data required, the effect of LoRA-adapting the CLIP text encoder while keeping T5 frozen, and the trade-off between a unified multi-task adapter and separate task-specific adapters. Together, these analyses provide practical guidance for parameter-efficient, few-shot adaptation of editing models to restoration.

\begin{figure*}[t]
  \centering
  \includegraphics[width=\textwidth]{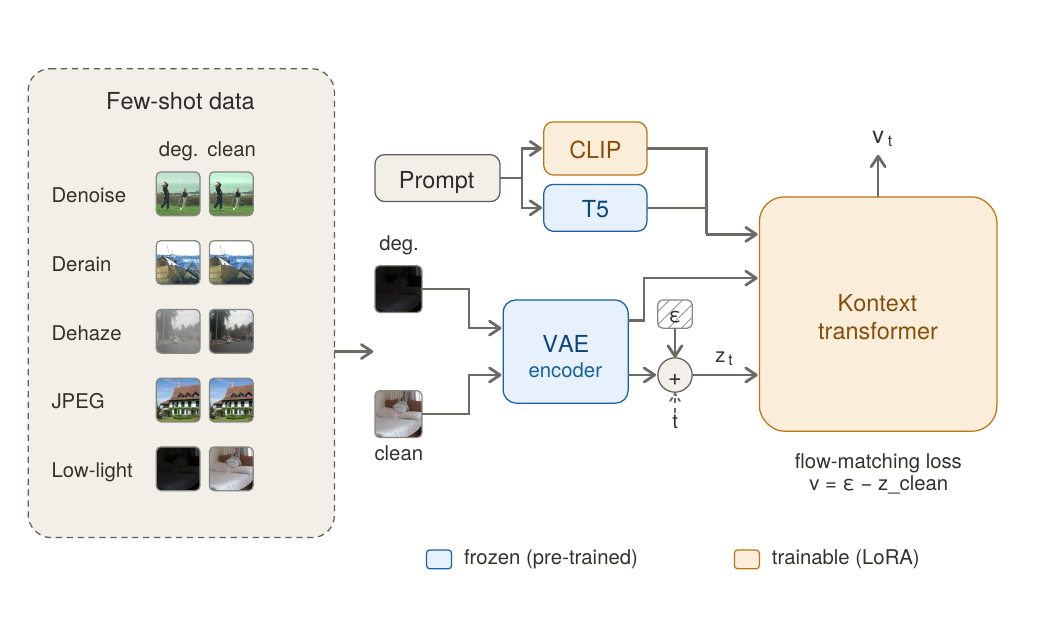}%
  \caption{Overview of the Edit2Restore pipeline. From a small few-shot set spanning five degradations, we sample a degraded--clean pair and its corresponding prompt and adapt the pre-trained FLUX.1 Kontext editing model with LoRA. A single adapter handles all five tasks, distinguished only by the text prompt.}
  \label{fig:method}
\end{figure*}

\section{Edit2Restore: Proposed Methodology}
\label{sec:method}

We propose a PEFT approach for adapting a pre-trained image editing model to text-conditioned multi-task image restoration with minimal training data. Our method leverages FLUX.1 Kontext~\cite{fluxkontext2025}, a 12-billion-parameter flow matching model designed for text-conditioned image-to-image translation, and adapts it to multi-task image restoration through LoRA. Figure~\ref{fig:method} illustrates our overall framework.

\subsection{Preliminaries: FLUX.1 Kontext}

FLUX.1 Kontext is a rectified flow transformer that performs image-to-image translation conditioned on both text instructions and input images. Given a degraded image $\mathbf{x}_{\text{deg}} \in \mathbb{R}^{H \times W \times 3}$ and a text prompt $\mathbf{t}$ describing the desired restoration operation, the model generates a restored image $\mathbf{x}_{\text{res}}$ through an iterative flow matching process in latent space.

The model consists of three main components that work together to generate images. A variational autoencoder with 16 latent channels encodes images into latent representations $\mathbf{z} = \mathcal{E}(\mathbf{x})$ and decodes them back to pixel space $\mathbf{x} = \mathcal{D}(\mathbf{z})$. Two text encoders, CLIP~\cite{clip} and T5~\cite{T5}, process the text prompt $\mathbf{t}$ into embeddings that condition the generation process. Finally, a flow matching transformer with a hybrid architecture combining double-stream and single-stream transformer blocks operates on the latent space, performing iterative refinement guided by text conditioning.

Unlike traditional diffusion models that predict noise, FLUX.1 Kontext uses rectified flow matching~\cite{rectifiedflow2023,lipman2023flow} which learns to predict velocity fields. The forward process constructs a linear interpolation path between clean latent $\mathbf{z}_{\text{clean}}$ and Gaussian noise $\epsilon \sim \mathcal{N}(0, \mathbf{I})$:
\begin{equation}
\mathbf{z}_t = (1 - t)\mathbf{z}_{\text{clean}} + t\epsilon
\label{eq:rectified_flow_forward}
\end{equation}
where $t \in [0, 1]$ is the timestep. The model learns to predict the velocity $\mathbf{v} = \epsilon - \mathbf{z}_{\text{clean}}$ that connects the two endpoints. This formulation provides straighter generation trajectories compared to standard diffusion processes, enabling faster and more stable sampling.

FLUX.1 Kontext is pre-trained on a large number of image-text pairs, learning rich priors about image structure, quality, and editing operations. Our key insight is that these pre-trained representations already encode knowledge about degradations and can be efficiently adapted for restoration.

\subsection{LoRA for Image Restoration}

Rather than fine-tuning all 12 billion parameters of FLUX.1 Kontext, we employ Low-Rank Adaptation to inject trainable parameters efficiently. For a pre-trained weight matrix $\mathbf{W}_0 \in \mathbb{R}^{d \times k}$ in the transformer layers, LoRA constrains the weight update through low-rank decomposition:
\begin{equation}
\mathbf{W} = \mathbf{W}_0 + \Delta\mathbf{W} = \mathbf{W}_0 + \mathbf{B}\mathbf{A}
\end{equation}
where $\mathbf{B} \in \mathbb{R}^{d \times r}$ and $\mathbf{A} \in \mathbb{R}^{r \times k}$ with rank $r \ll \min(d, k)$. During training, $\mathbf{W}_0$ remains frozen while $\mathbf{B}$ and $\mathbf{A}$ are trained.

We apply LoRA to the attention projection layers in the flow matching transformer. Specifically, we inject LoRA adapters into the query, key, and value projection matrices of each attention block. This results in a small fraction of trainable parameters, while maintaining the model's expressive capacity for the restoration task.

\subsection{Text Prompt Design}

Rather than a single rigid template, we condition each restoration task on a concise natural-language instruction that specifies both the degradation to remove and the desired clean outcome. Table~\ref{tab:prompts} lists the prompt used for each of the five tasks.

Our prompt design keeps instructions short and human-readable, which serves two purposes: \textbf{(1)} it leverages FLUX.1 Kontext's pre-trained understanding of editing instructions without requiring extensive prompt engineering, and \textbf{(2)} it enables straightforward multi-task learning, where the model distinguishes tasks primarily through the prompt conditioning.  For the unified multi-task setting, we condition each training sample on its corresponding task-specific prompt, allowing a single LoRA adapter to handle all five degradation types. During inference, the appropriate prompt is provided to guide the restoration process.

\begin{table}[t]
\centering
\caption{Text prompts used for each restoration task, applied at both LoRA training and inference time. We use these prompts for our method and all editing-model baselines, except RealRestorer~\cite{realrestorer}, which relies on its own per-task prompts.}
\label{tab:prompts}
\setlength{\tabcolsep}{4pt}
\renewcommand{\arraystretch}{1.3}
\resizebox{0.95\textwidth}{!}{%
\begin{tabular}{@{}l l@{}}
\toprule
Task & Text prompt \\
\midrule
Denoising & \textit{Remove noise and grain, restore a clean and sharp image} \\
Deraining & \textit{Remove rain streaks and restore the clean image} \\
Dehazing & \textit{Remove haze and fog, restore clear visibility and natural colors} \\
Compression artifact removal & \textit{Remove compression artifacts and restore the clean image} \\
Low-light enhancement & \textit{Brighten this photo, recover shadow details, reduce noise, keep colors natural} \\
\bottomrule
\end{tabular}}
\end{table}

\section{Experiments}
We present comprehensive experiments analyzing our parameter-efficient adaptation approach across five restoration tasks. We first compare against state-of-the-art editing models in the zero-shot regime and a strong large-data restoration baseline, then examine the key design choices governing our adaptation: the effect of text encoder fine-tuning, the amount of paired data required, and the trade-off between unified multi-task and task-specific adapters.

\subsection{Experimental Setup}

\noindent\textbf{Datasets.} Rather than training on full restoration datasets, we construct small few-shot training sets by sampling a limited number of clean images per task, using 32, 64, and 128 samples to characterize the few-shot regime. \textit{Denoising:} we sample clean images from BSD400~\cite{bsd400} and synthesize Gaussian noise at $\sigma \in \{15, 25, 50\}$ in approximately equal proportions, evaluating on BSD68~\cite{bsd68} at the same three levels. \textit{Deraining:} we sample degraded--clean pairs from Rain100L~\cite{rain100l} and evaluate on its test split. \textit{Dehazing:} we sample degraded--clean pairs from the RESIDE Outdoor Training Set (OTS)~\cite{sots} and evaluate on the standard outdoor split of the Synthetic Objective Testing Set (SOTS), containing 500 images. \textit{JPEG compression artifact removal:} we sample clean images from DIV2K~\cite{div2k} and synthesize JPEG compression at quality factors $q \in \{1, 10, 20\}$ in approximately equal proportions, evaluating on the 24 Kodak images at the same factors. \textit{Low-light enhancement:} we use the real low-/normal-light pairs of LOL v1~\cite{lol}, sampling from its training split and evaluating on its 15 test images. These benchmarks provide diverse test conditions to assess generalization beyond the limited training samples.

\noindent\textbf{Image resolution handling.} During training, all samples are resized to $1024 \times 1024$ resolution. At test time, input images are first resized to $1024 \times 1024$ for processing, and the restored output is then resized back to the original resolution, ensuring compatibility with arbitrary input dimensions while maintaining the model's expected operating resolution.

\noindent\textbf{Implementation details.} All experiments are conducted on a single NVIDIA H100 GPU. We adapt FLUX.1 Kontext with LoRA while keeping the transformer backbone, the T5 text encoder, and the VAE frozen. In the transformer, LoRA is injected into the attention projection layers of every block and on the text side, we LoRA-adapt only the CLIP encoder, again at its attention projections. The LoRA rank and batch size are set to $64$ and $4$, respectively, for all experiments. For each task, we experiment with three training-set sizes: 32, 64, and 128 sampled images to characterize performance in the few-shot regime. Each training run is performed for 1,920 iterations across all experimental configurations to ensure fair comparison. We employ the AdamW optimizer~\cite{adamw} with learning rate $1 \times 10^{-4}$ for the transformer LoRA parameters. For the CLIP LoRA parameters, we use a lower learning rate of $5 \times 10^{-6}$ to prevent catastrophic forgetting of the pre-trained text representations. Weight decay is set to $1 \times 10^{-4}$ for the transformer parameters and $1 \times 10^{-3}$ for the text encoder parameters. We apply a constant learning rate schedule with 500 warmup steps. Training is performed in mixed precision (bfloat16) with gradient checkpointing enabled to reduce memory consumption.

During training, we use logit-normal timestep sampling with $\mu = 0.0$ and $\sigma = 1.0$ for the rectified flow matching objective. Dataset entries are shuffled before training to mix all five degradation types uniformly across batches. During inference, we follow the rectified flow sampling procedure. Starting from pure noise, we iteratively integrate the learned velocity field to generate the clean latent. For all experiments, we use 28 integration steps and set the guidance scale to 2.5 following FLUX.1 Kontext's default sampling schedule. Inference takes approximately 15-20 seconds per image at $1024 \times 1024$ resolution on a single H100 GPU.

\noindent\textbf{Evaluation metrics.} We report three complementary perceptual metrics, all lower-is-better, that assess restoration quality from different perspectives: (1) \textbf{LPIPS (Learned Perceptual Image Patch Similarity)}~\cite{lpips} measures full-reference perceptual similarity between each restored image and its ground truth using deep network features, correlating well with human judgments of visual similarity; (2) \textbf{FID (Fréchet Inception Distance)}~\cite{fid} measures distribution-level similarity between the sets of restored and ground-truth images, capturing overall perceptual quality and realism; and (3) \textbf{CMMD (CLIP Maximum Mean Discrepancy)}~\cite{cmmd} evaluates distribution matching in CLIP feature space via maximum mean discrepancy, providing a robust alternative to FID. Together, LPIPS captures per-image fidelity to the reference while FID and CMMD capture distributional realism. We deliberately favor these perceptual and distribution-level metrics over pixel-fidelity scores (PSNR/SSIM), consistent with our approach's focus on leveraging pre-trained generative priors to produce perceptually plausible restorations rather than minimizing pixel-wise error.

\begin{table}[t]
\centering
\caption{Quantitative comparison on denoising (BSD68), deraining (Rain100L), and dehazing (SOTS outdoor). Lower values are better for all metrics ($\downarrow$): \textbf{L}=LPIPS, \textbf{F}=FID, and \textbf{C}=CMMD. Best and second-best values per column are shown in bold and underlined, respectively. $^{\dagger}$ denotes no LoRA adaptation of CLIP; T5 is frozen in all variants.}
\label{tab:main_classic}
\setlength{\tabcolsep}{4pt}
\resizebox{\textwidth}{!}{
\begin{tabular}{lccccccccccccccc}
\toprule
\multirow{2}{*}{Method} & \multicolumn{3}{c}{Denoise $\sigma$15} & \multicolumn{3}{c}{Denoise $\sigma$25} & \multicolumn{3}{c}{Denoise $\sigma$50} & \multicolumn{3}{c}{Deraining} & \multicolumn{3}{c}{Dehazing} \\
\cmidrule(lr){2-4} \cmidrule(lr){5-7} \cmidrule(lr){8-10} \cmidrule(lr){11-13} \cmidrule(lr){14-16}
 & L$\downarrow$ & F$\downarrow$ & C$\downarrow$ & L$\downarrow$ & F$\downarrow$ & C$\downarrow$ & L$\downarrow$ & F$\downarrow$ & C$\downarrow$ & L$\downarrow$ & F$\downarrow$ & C$\downarrow$ & L$\downarrow$ & F$\downarrow$ & C$\downarrow$ \\
\midrule
FLUX.1 Kontext& 0.105 & 33.1 & 0.776 & 0.141 & 41.2 & 0.820 & 0.237 & 73.8 & 0.963 & 0.067 & 23.6 & 0.537 & 0.219 & 27.7 & 0.536 \\
Qwen-Image-Edit (2509) & \underline{0.096} & \textbf{24.8} & \textbf{0.413} & \textbf{0.119} & \textbf{33.0} & \textbf{0.432} & 0.222 & 59.3 & \textbf{0.428} & 0.120 & 31.1 & 0.597 & 0.202 & 28.8 & 0.447 \\
Qwen-Image-Edit (2511) & 0.126 & 33.9 & 0.794 & 0.158 & 43.3 & 0.875 & 0.232 & 64.0 & 1.004 & 0.134 & 37.3 & 0.826 & 0.222 & 30.8 & 0.690 \\
Step1X-Edit & 0.374 & 94.0 & 1.208 & 0.446 & 103.9 & 1.351 & 0.585 & 145.3 & 1.865 & 0.180 & 55.5 & 0.975 & 0.547 & 95.4 & 1.774 \\
FLUX.2 Klein-9B & 0.202 & 49.4 & 1.020 & 0.229 & 57.3 & 1.028 & 0.305 & 80.9 & 1.078 & 0.117 & 34.2 & 0.696 & 0.268 & 38.6 & 1.042 \\
LongCat-Edit & 0.229 & 49.3 & 0.644 & 0.267 & 64.1 & 0.695 & 0.354 & 89.2 & 0.902 & 0.139 & 39.7 & 0.709 & 0.313 & 38.3 & 0.943 \\
RealRestorer & 0.127 & 44.6 & 0.854 & 0.157 & 52.0 & 0.810 & 0.243 & 67.4 & 0.846 & 0.089 & 27.4 & 0.761 & 0.150 & 24.9 & 0.370 \\
\midrule
\textbf{Ours} (Unified, 32) & \textbf{0.095} & \underline{28.0} & \underline{0.554} & \underline{0.125} & \underline{35.6} & \underline{0.597} & \textbf{0.187} & 54.8 & \underline{0.729} & \underline{0.031} & \underline{11.8} & 0.411 & \underline{0.082} & \underline{16.6} & \textbf{0.118} \\
\textbf{Ours} (Unified, 64) & 0.101 & 30.7 & 0.568 & 0.133 & 38.1 & 0.626 & 0.189 & \textbf{52.7} & 0.745 & \underline{0.031} & \underline{11.8} & \underline{0.403} & 0.089 & 17.5 & 0.132 \\
\textbf{Ours} (Unified, 128) & 0.101 & 30.2 & 0.601 & 0.133 & 37.8 & 0.647 & 0.193 & 53.3 & 0.782 & 0.032 & 11.9 & 0.408 & 0.084 & 16.8 & \underline{0.120} \\
\textbf{Ours} (Unified, 128)$^{\dagger}$ & 0.104 & 30.6 & 0.650 & 0.135 & 38.8 & 0.714 & 0.201 & 60.1 & 0.887 & 0.036 & 12.7 & 0.430 & 0.094 & 18.5 & 0.131 \\
\textbf{Ours} (Task-specific, 128) & 0.111 & 34.8 & 0.664 & 0.141 & 41.3 & 0.719 & \underline{0.188} & \underline{53.0} & 0.815 & \textbf{0.030} & \textbf{11.4} & \textbf{0.394} & \textbf{0.065} & \textbf{14.4} & 0.123 \\
\bottomrule
\end{tabular}}
\end{table}

\begin{table}[t]
\centering
\caption{Quantitative comparison on low-light enhancement (LOL) and JPEG compression restoration (Kodak). Lower values are better for all metrics ($\downarrow$): \textbf{L}=LPIPS, \textbf{F}=FID, and \textbf{C}=CMMD. Best and second-best values per column are shown in bold and underlined, respectively. $^{\dagger}$ denotes no LoRA adaptation of CLIP; T5 is frozen in all variants.}
\label{tab:main_lowlight_compression}
\setlength{\tabcolsep}{4pt}
\resizebox{\textwidth}{!}{
\begin{tabular}{lcccccccccccc}
\toprule
\multirow{2}{*}{Method} & \multicolumn{3}{c}{Low-light} & \multicolumn{3}{c}{JPEG $q$=1} & \multicolumn{3}{c}{JPEG $q$=10} & \multicolumn{3}{c}{JPEG $q$=20} \\
\cmidrule(lr){2-4} \cmidrule(lr){5-7} \cmidrule(lr){8-10} \cmidrule(lr){11-13}
 & L$\downarrow$ & F$\downarrow$ & C$\downarrow$ & L$\downarrow$ & F$\downarrow$ & C$\downarrow$ & L$\downarrow$ & F$\downarrow$ & C$\downarrow$ & L$\downarrow$ & F$\downarrow$ & C$\downarrow$ \\
\midrule
FLUX.1 Kontext & 0.280 & 84.4 & 0.453 & 0.575 & 295.4 & 2.809 & 0.289 & 73.3 & 0.917 & 0.182 & 37.5 & 0.722 \\
Qwen-Image-Edit (2509) & 0.394 & 124.4 & 0.933 & 0.537 & 147.3 & \underline{0.947} & 0.302 & 55.4 & 0.626 & 0.255 & 37.2 & 0.492 \\
Qwen-Image-Edit (2511)& 0.322 & 101.6 & 0.745 & 0.494 & 142.9 & 1.079 & 0.266 & 51.8 & 0.663 & 0.212 & 35.3 & 0.548 \\
Step1X-Edit & 0.351 & 126.5 & 0.584 & 0.624 & 214.6 & 2.363 & 0.395 & 89.2 & 0.958 & 0.311 & 76.5 & 0.844 \\
FLUX.2 Klein-9B & 0.257 & 84.4 & 0.556 & 0.541 & 161.3 & 1.417 & 0.295 & 62.8 & 0.779 & 0.243 & 48.6 & 0.669 \\
LongCat-Edit & 0.316 & 112.7 & 0.324 & 0.592 & 262.0 & 2.086 & 0.329 & 70.6 & 0.661 & 0.230 & 48.6 & 0.525 \\
RealRestorer & 0.229 & 84.2 & 0.371 & 0.495 & 144.8 & 1.165 & 0.279 & 57.3 & 0.662 & 0.233 & 43.8 & 0.600 \\
\midrule
\textbf{Ours} (Unified, 32) & \textbf{0.140} & \textbf{57.3} & \textbf{0.220} & \textbf{0.393} & \textbf{126.5} & \textbf{0.846} & \textbf{0.208} & \textbf{42.0} & \textbf{0.370} & \textbf{0.161} & \textbf{27.1} & \textbf{0.328} \\
\textbf{Ours} (Unified, 64) & 0.165 & 59.8 & 0.288 & 0.415 & 132.8 & 1.032 & 0.217 & 43.7 & 0.406 & 0.165 & \underline{29.0} & 0.354 \\
\textbf{Ours} (Unified, 128) & \underline{0.144} & \underline{58.0} & \underline{0.235} & \underline{0.414} & \underline{128.9} & 1.045 & \underline{0.213} & \underline{42.6} & \underline{0.391} & \underline{0.163} & \underline{29.0}  & 0.347 \\
\textbf{Ours} (Unified, 128)$^{\dagger}$ & 0.177 & 66.9 & 0.317 & 0.464 & 167.3 & 1.341 & 0.251 & 50.0 & 0.549 & 0.196 & 34.4 & 0.529 \\
\textbf{Ours} (Task-specific, 128) & 0.148 & 61.1 & 0.243 & 0.431 & 130.6 & 1.245 & 0.222 & 46.0 & 0.416 & 0.168 & 30.2 & \underline{0.345} \\
\bottomrule
\end{tabular}}
\end{table}

\begin{figure}[!t]
  \centering
\includegraphics[height=0.75\textheight, width=0.90\textwidth]{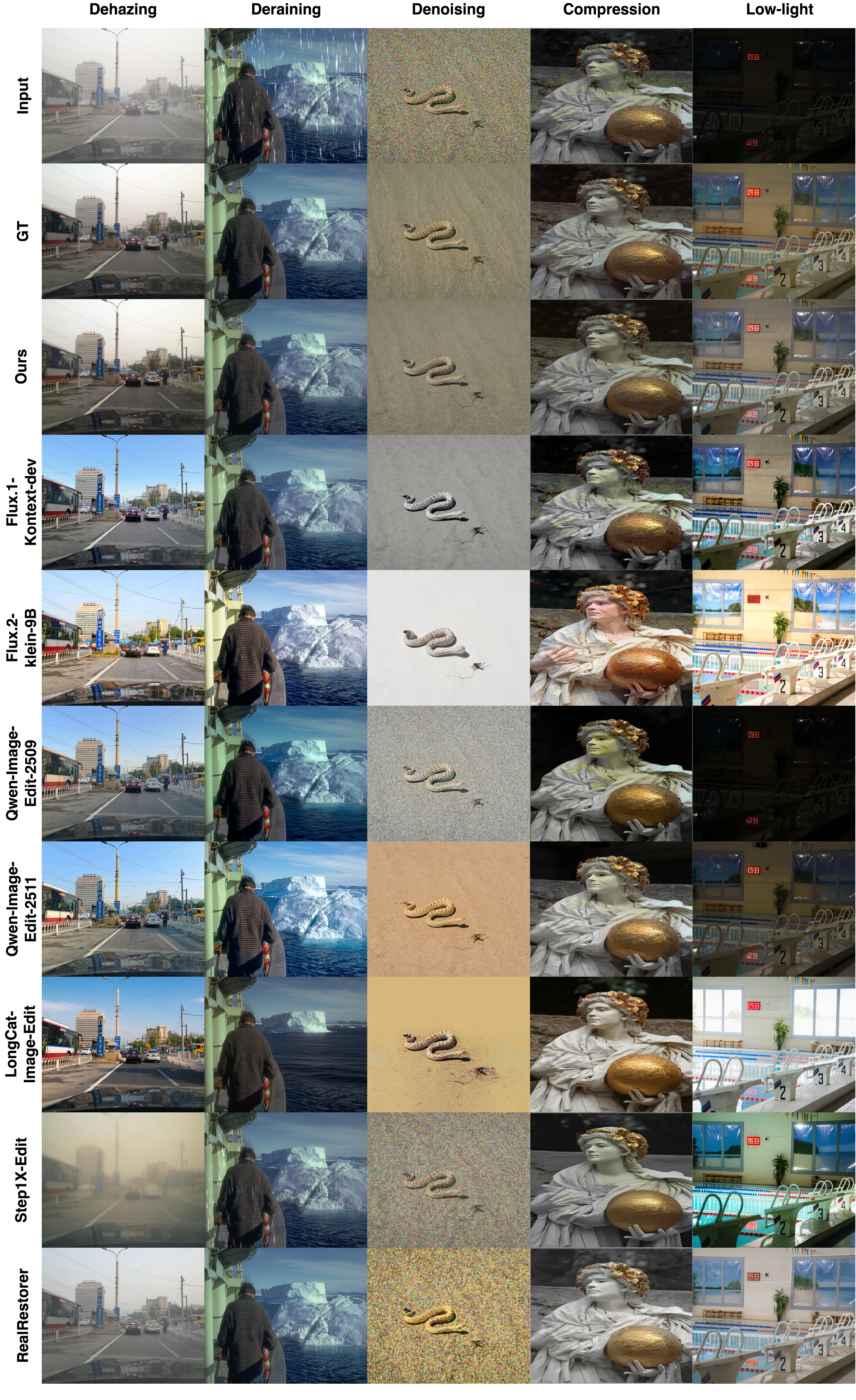}
  \caption{Qualitative comparison across the five restoration tasks (columns). From top to bottom: degraded input, ground truth, our few-shot adapted model (Unified, 32 samples per task), and the baselines. Our model removes the degradation while preserving structure, texture, and color, most closely matching the ground truth.}
  \label{fig:qualitative_comparison}
\end{figure}

\subsection{Comparison with State-of-the-Art}

We benchmark our approach against two families of methods: (i)~state-of-the-art text-conditioned editing models evaluated in the zero-shot regime (FLUX.1 Kontext, Qwen-Image-Edit 2509/2511, Step1X-Edit, FLUX.2 Klein-9B, LongCat-Edit), all prompted with the instructions in Table~\ref{tab:prompts}, and (ii)~RealRestorer, a recent restoration model that fine-tunes a large editing backbone on more than one million curated pairs. Tables~\ref{tab:main_classic} and~\ref{tab:main_lowlight_compression} report results across all five tasks, and Figure~\ref{fig:qualitative_comparison} shows representative qualitative comparisons.

\noindent\textbf{Zero-shot editing baselines.} Despite their strong generative priors, none of the editing models reliably restores degraded inputs without adaptation. This is most telling for FLUX.1 Kontext, the very backbone we fine-tune: in its zero-shot form it is only a mediocre restorer, struggling especially as degradations become severe (e.g., its FID exceeds 290 at the harshest JPEG setting and rises sharply with noise level). Step1X-Edit is the weakest by a wide margin, and FLUX.2 Klein-9B and LongCat-Edit also degrade sharply on heavy corruptions, both exceeding an FID of 160 at the most severe JPEG setting. The one clear exception is Qwen-Image-Edit (2509), a strong zero-shot restorer on low-noise denoising: it achieves the best FID and CMMD at $\sigma{=}15$ and the best scores across all three metrics at $\sigma{=}25$, while our Unified-32 model achieves the best LPIPS at $\sigma{=}15$. This strength, however, is largely confined to mild Gaussian noise, and on the other degradations Qwen falls well behind our adapted model. That the same FLUX.1 Kontext backbone moves from mediocre zero-shot restoration to the strongest overall performance among the evaluated methods once adapted, while strong editors like Qwen remain limited without adaptation, supports our central premise: raw editing capacity does not, on its own, confer reliable restoration ability; targeted adaptation is what unlocks it.

\noindent\textbf{Comparison with RealRestorer.} RealRestorer is the most directly comparable baseline, as it is the only competing method explicitly designed and fine-tuned for restoration, trained on a vast number of curated pairs. Even so, our few-shot model surpasses it on the large majority of metrics while using three to four orders of magnitude less data. The margins are widest precisely where editing priors are least directly useful: on low-light and dehazing we improve its FID and CMMD by a large margin. RealRestorer stays competitive only on a handful of denoising entries, where it is in turn outperformed by Qwen-Image-Edit family.

\noindent\textbf{Per-task analysis.} The task-by-task picture in Tables~\ref{tab:main_classic} and~\ref{tab:main_lowlight_compression} is consistent with this summary. Denoising is the one task where a zero-shot editor remains competitive: Qwen-Image-Edit (2509) is the strongest method at low and moderate noise, while our Unified-32 model is the steady runner-up and already gives the best perceptual similarity at $\sigma{=}15$. The ranking shifts at high noise, where our adapters take the best LPIPS and FID at $\sigma{=}50$, even if Qwen retains the lowest CMMD throughout. On deraining and dehazing the picture is unambiguous: our models lead every metric, with the task-specific adapter best overall and the unified variants close behind, roughly halving the FID of the strongest baseline (zero-shot FLUX.1 Kontext on deraining, RealRestorer on dehazing). For compression artifact removal, our Unified-32 model is best at every quality factor and the gap widens as compression becomes more severe, where most baselines collapse while ours degrades gracefully. For low-light, our smallest model again leads all three metrics; RealRestorer is the best baseline but trails by a clear margin, and the Qwen models perform poorly. Overall, our Unified-32 model is frequently the single best entry, demonstrating that competitive restoration can emerge from parameter-efficient fine-tuning of a strong editing backbone with very little task-specific data.

\noindent\textbf{Qualitative comparison.} Figure~\ref{fig:qualitative_comparison} makes these differences visible and exposes two recurring failure modes among the baselines: under-restoration, where the degradation is only partially removed, and over-editing, where the degradation is removed but the content, contrast, or color is altered. In dehazing, Step1X-Edit and RealRestorer leave a visible haze, while the remaining editors over-dehaze and shift the contrast and color away from the reference; our result is the closest to the ground truth and preserves the original colors. Deraining shows the same split: several methods (FLUX.1 Kontext, Step1X-Edit, RealRestorer) leave residual rain streaks, and others remove the rain but distort contrast and color, whereas ours removes the streaks while keeping the scene faithful. Denoising (at $\sigma{=}50$) is the most revealing case, as some baselines retain clear noise patterns while others suppress the noise only by rewriting the image content; our model is the most reliable, removing the noise while preserving the snake's scales and the surrounding texture. Under compression, some methods leave blocking artifacts and others alter the statue and background entirely, while ours removes the artifacts and keeps the subject faithful to the reference. Finally, low-light is the most challenging: some baselines barely brighten the scene, some over-expose it, and others raise the brightness but badly distort the color distribution; our output recovers a natural exposure and color balance closest to the ground truth. These observations mirror the quantitative results: among the evaluated methods, ours most consistently removes the degradation while preserving content fidelity.

\subsection{Effect of Text Encoder Adaptation}

FLUX.1 Kontext conditions generation on two pre-trained text encoders, CLIP and T5, that map the restoration prompt into the embedding space steering the transformer. Since these encoders were trained for general-purpose editing, it is not obvious whether they already supply adequate conditioning for restoration or whether they should be adapted alongside the transformer. We isolate this factor by repeating the unified 128-sample run with the encoders held frozen and everything else unchanged; in the adapted variant we attach LoRA to the CLIP encoder while T5 remains frozen, following the same low-rank scheme used for the transformer. The frozen variant is marked $^{\dagger}$ in Tables~\ref{tab:main_classic} and~\ref{tab:main_lowlight_compression}.

Adapting CLIP helps on every task, and the magnitude of the gain varies in an informative way. On denoising, deraining, and dehazing the improvement is steady but modest: FID falls from 60.1 to 53.3 at $\sigma{=}50$, from 12.7 to 11.9 on deraining, and from 18.5 to 16.8 on dehazing, with CMMD tracking the same trend (for instance $0.887\!\rightarrow\!0.782$ at $\sigma{=}50$). The benefit is larger on low-light enhancement and severe JPEG restoration, where low-light FID improves from 66.9 to 58.0 and JPEG at $q{=}1$ from 167.3 to 128.9. On these tasks the restored output departs furthest from the input, in exposure and color or in detail lost to heavy quantization, and adapting the conditioning representation rather than freezing it is what closes the gap. Given that the improvement is uniform across tasks and its cost is small next to the transformer adapters, we LoRA-adapt the CLIP encoder in all remaining experiments.

\subsection{Data Efficiency Analysis}

The central claim of our work is that adaptation needs very little supervision, so the natural question is how restoration quality scales with the size of the training set. We train the unified adapter at 32, 64, and 128 samples per task, corresponding to 160, 320, and 640 images in total across the five degradations, with everything else held fixed; all configurations appear in Tables~\ref{tab:main_classic} and~\ref{tab:main_lowlight_compression}.

Quality is essentially flat across this range, and the small differences that do appear are non-monotonic rather than improving with scale. Dehazing FID stays within 16.6--17.5, deraining is effectively constant at 11.8--11.9, and low-light remains within 57.3--59.8. In several columns the smallest, 32-sample configuration is in fact the best, including dehazing, low-light, and JPEG at $q{=}10$, while the 64- and 128-sample variants take the lead only on isolated denoising levels (for instance $\sigma{=}50$). The lack of any consistent gain from added data suggests that the visual priors in FLUX.1 Kontext already carry much of what restoration requires, and that 32 paired images per task are sufficient to achieve near-saturated performance over the tested 32--128 sample range, with rapidly diminishing returns from additional examples.

\subsection{Unified Multi-Task vs Task-Specific Adapters}

A natural objection to a single multi-task adapter is that conditioning one set of weights on five different prompts dilutes its capacity, and that an adapter trained for one degradation should restore it better. We test this directly by comparing the unified model against adapters trained independently per task, both at 128 samples, in Tables~\ref{tab:main_classic} and~\ref{tab:main_lowlight_compression}.

On deraining and dehazing, the task-specific adapter has a modest edge, as expected when a dedicated set of weights can specialize to a single, well-defined corruption. On every other task the unified adapter is equal or better: it leads on denoising at $\sigma{=}15$ and $\sigma{=}25$, and on low-light and all three compression levels. More striking still, the unified adapter trained with only 32 samples per task frequently matches or beats the task-specific adapter trained with 128, despite using a quarter of the data and a single shared set of weights.

The practical case is just as strong. One unified adapter covers all five degradations, cutting storage roughly five-fold and removing any need to swap checkpoints at inference. Taken together, accuracy and deployment cost point the same way: for few-shot restoration, unified multi-task adaptation should be the default, and specialized per-task models are not the prerequisite for strong perceptual quality.

\section{Conclusion}

We have shown that a pre-trained text-conditioned image editing model can be turned into a competitive image restorer with remarkably little supervision. FLUX.1 Kontext and other state-of-the-art editors restore poorly in the zero-shot setting, but parameter-efficient LoRA adaptation with only tens of paired images per task unlocks substantial gains across five degradations. The resulting few-shot model surpasses RealRestorer, a restoration baseline trained on more than one million curated pairs, on the large majority of metrics while using three to four orders of magnitude less data.

Three findings emerge from our ablations. Restoration quality is essentially flat across the tested training-set sizes, with 32 paired images per task sufficient to achieve near-saturated performance over the 32--128 sample range and little consistent gain from additional examples, suggesting that the priors learned during large-scale editing pre-training already transfer much of what restoration requires. LoRA-adapting the CLIP encoder alongside the transformer adapters helps on every task and most on the cases where input and output differ most, showing that the conditioning pathway, not only the visual one, benefits from adaptation. And a single unified adapter matches or exceeds per-task adapters in the few-shot regime while cutting storage roughly five-fold. By achieving this with orders of magnitude fewer examples than both from-scratch restoration networks and recent large-data editing-based methods, our work offers a practical route to restoration in data-scarce settings. 

\clearpage

 \section*{Acknowledgements}
This work was supported by Codeway. A.M. Tekalp acknowledges support from Turkish Academy of Sciences (T\"UBA).

%
%
\bibliographystyle{splncs04}
\bibliography{main}

@String(IJCV  = {Int. J. Comput. Vis.})

@String(CVPR  = {IEEE Conf. Comput. Vis. Pattern Recog.})

@String(ICCV  = {Int. Conf. Comput. Vis.})

@String(ECCV  = {Eur. Conf. Comput. Vis.})

@String(NeurIPS = {Adv. Neural Inform. Process. Syst.})

@String(ICML  = {Int. Conf. Mach. Learn.})

@String(ICLR  = {Int. Conf. Learn. Represent.})

@String(BMVC  = {Brit. Mach. Vis. Conf.})

@String(CVPRW = {IEEE Conf. Comput. Vis. Pattern Recog. Worksh.})

@String(AAAI  = {AAAI})

@String(JMLR  = {J. Mach. Learn. Res.})

@String(TIP   = {IEEE Trans. Image Process.})

@String(IJCV  = {IJCV})

@String(CVPR  = {CVPR})

@String(ICCV  = {ICCV})

@String(ECCV  = {ECCV})

@String(NeurIPS = {NeurIPS})

@String(ICML  = {ICML})

@String(ICLR  = {ICLR})

@String(BMVC  =	{BMVC})

@String(CVPRW = {CVPRW})

@String(JMLR  = {JMLR})

@String(TIP   = {IEEE TIP})

@inproceedings{lora2022,
  title={{LoRA}: Low-Rank Adaptation of Large Language Models},
  author={Hu, Edward J. and Shen, Yelong and Wallis, Phillip and Allen-Zhu, Zeyuan and Li, Yuanzhi and Wang, Shean and Wang, Lu and Chen, Weizhu},
  booktitle=ICLR,
  year={2022}
}

@article{fluxkontext2025,
  title={{FLUX.1 Kontext}: Flow Matching for In-Context Image Generation and Editing in Latent Space},
  author={{Black Forest Labs} and Batifol, Stephen and Blattmann, Andreas and Boesel, Frederic and others},
  journal={arXiv preprint arXiv:2506.15742},
  year={2025}
}

@inproceedings{promptir2023,
  title={{PromptIR}: Prompting for All-in-One Blind Image Restoration},
  author={Potlapalli, Vaishnav and Zamir, Syed Waqas and Khan, Salman and Khan, Fahad Shahbaz},
  booktitle=NeurIPS,
  year={2023}
}

@inproceedings{instructir2024,
  title={{InstructIR}: High-Quality Image Restoration Following Human Instructions},
  author={Conde, Marcos V. and Geigle, Gregor and Timofte, Radu},
  booktitle=ECCV,
  year={2024}
}

@inproceedings{swinir2021,
  title={{SwinIR}: Image Restoration Using Swin Transformer},
  author={Liang, Jingyun and Cao, Jiezhang and Sun, Guolei and Zhang, Kai and Van Gool, Luc and Timofte, Radu},
  booktitle={IEEE Int. Conf. Comput. Vis. Workshops},
  year={2021},
  pages={1833--1844}
}

@inproceedings{restormer2022,
  title={Restormer: Efficient Transformer for High-Resolution Image Restoration},
  author={Zamir, Syed Waqas and Arora, Aditya and Khan, Salman and Hayat, Munawar and Khan, Fahad Shahbaz and Yang, Ming-Hsuan},
  booktitle=CVPR,
  year={2022},
  pages={5728--5739}
}

@inproceedings{airnet2022,
  title={All-in-One Image Restoration for Unknown Corruption},
  author={Li, Boyun and Liu, Xiao and Hu, Peng and Wu, Zhongqin and Lv, Jiancheng and Peng, Xi},
  booktitle=CVPR,
  year={2022},
  pages={17452--17462}
}

@inproceedings{rectifiedflow2023,
  title={Flow Straight and Fast: Learning to Generate and Transfer Data with Rectified Flow},
  author={Liu, Xingchao and Gong, Chengyue and Liu, Qiang},
  booktitle=ICLR,
  year={2023}
}

@inproceedings{lipman2023flow,
  title={Flow Matching for Generative Modeling},
  author={Lipman, Yaron and Chen, Ricky T. Q. and Ben-Hamu, Heli and Nickel, Maximilian and Le, Matthew},
  booktitle=ICLR,
  year={2023}
}

@inproceedings{
adamw,
title={Decoupled Weight Decay Regularization},
author={Ilya Loshchilov and Frank Hutter},
booktitle={International Conference on Learning Representations},
year={2019},
url={https://openreview.net/forum?id=Bkg6RiCqY7},
}

@INPROCEEDINGS{bsd68,
  author={Martin, D. and Fowlkes, C. and Tal, D. and Malik, J.},
  booktitle=ICCV, 
  title={A database of human segmented natural images and its application to evaluating segmentation algorithms and measuring ecological statistics}, 
  year={2001},
  volume={2},
  number={},
  pages={416-423 vol.2},
  doi={10.1109/ICCV.2001.937655}}

@ARTICLE{bsd400,
  author={Arbeláez, Pablo and Maire, Michael and Fowlkes, Charless and Malik, Jitendra},
  journal={IEEE Transactions on Pattern Analysis and Machine Intelligence}, 
  title={Contour Detection and Hierarchical Image Segmentation}, 
  year={2011},
  volume={33},
  number={5},
  pages={898-916},
  doi={10.1109/TPAMI.2010.161}}

@ARTICLE{sots,
  author={Li, Boyi and Ren, Wenqi and Fu, Dengpan and Tao, Dacheng and Feng, Dan and Zeng, Wenjun and Wang, Zhangyang},
  journal=TIP, 
  title={Benchmarking Single-Image Dehazing and Beyond}, 
  year={2019},
  volume={28},
  number={1},
  pages={492-505},
  doi={10.1109/TIP.2018.2867951}}

@inproceedings{fid,
 author = {Heusel, Martin and Ramsauer, Hubert and Unterthiner, Thomas and Nessler, Bernhard and Hochreiter, Sepp},
 booktitle = {Advances in Neural Information Processing Systems},
 editor = {I. Guyon and U. Von Luxburg and S. Bengio and H. Wallach and R. Fergus and S. Vishwanathan and R. Garnett},
 pages = {},
 publisher = {Curran Associates, Inc.},
 title = {GANs Trained by a Two Time-Scale Update Rule Converge to a Local Nash Equilibrium},
 url = {https://proceedings.neurips.cc/paper_files/paper/2017/file/8a1d694707eb0fefe65871369074926d-Paper.pdf},
 volume = {30},
 year = {2017}
}

@INPROCEEDINGS{cmmd,
  author={Jayasumana, Sadeep and Ramalingam, Srikumar and Veit, Andreas and Glasner, Daniel and Chakrabarti, Ayan and Kumar, Sanjiv},
  booktitle=CVPR, 
  title={Rethinking FID: Towards a Better Evaluation Metric for Image Generation}, 
  year={2024},
  volume={},
  number={},
  pages={9307-9315},
  doi={10.1109/CVPR52733.2024.00889}}

@ARTICLE{DnCNN,
  author={Zhang, Kai and Zuo, Wangmeng and Chen, Yunjin and Meng, Deyu and Zhang, Lei},
  journal=TIP, 
  title={Beyond a Gaussian Denoiser: Residual Learning of Deep CNN for Image Denoising}, 
  year={2017},
  volume={26},
  number={7},
  pages={3142-3155},
  doi={10.1109/TIP.2017.2662206}
}

@InProceedings{GoPro,
  author = {Nah, Seungjun and Kim, Tae Hyun and Lee, Kyoung Mu},
  title = {Deep Multi-Scale Convolutional Neural Network for Dynamic Scene Deblurring},
  booktitle = CVPR,
  month = {July},
  year = {2017}
}

@article{DeblurGAN,
  title = {DeblurGAN: Blind Motion Deblurring Using Conditional Adversarial Networks},
  author = {Kupyn, Orest and Budzan, Volodymyr and Mykhailych, Mykola and Mishkin, Dmytro and Matas, Jiri},
  journal = {ArXiv e-prints},
  eprint = {1711.07064},
  year = 2017
}

@InProceedings{RetinexFormer,
    author    = {Cai, Yuanhao and Bian, Hao and Lin, Jing and Wang, Haoqian and Timofte, Radu and Zhang, Yulun},
    title     = {Retinexformer: One-stage Retinex-based Transformer for Low-light Image Enhancement},
    booktitle = ICCV,
    month     = {October},
    year      = {2023},
    pages     = {12504-12513}
}

@inproceedings{FFA-Net,
    title={FFA-Net: Feature fusion attention network for single image dehazing},
    author={Qin, Xu and Wang, Zhilin and Bai, Yuanchao and Xie, Xiaodong and Jia, Huizhu},
    booktitle=AAAI,
    volume={34},
    pages={11908--11915},
    year={2020}
}

@InProceedings{MSFPN,
	author = {Jiang, Kui and Wang, Zhongyuan and Yi, Peng and Chen, Chen and Huang, Baojin and Luo, Yimin and Ma, Jiayi and Jiang, Junjun},
	title = {Multi-Scale Progressive Fusion Network for Single Image Deraining},
	booktitle = CVPR,
	month = {June},
	year = {2020}
}

@article{DesnowNet,
  title={DesnowNet: Context-Aware Deep Network for Snow Removal},
  author={Liu, Yun-Fu and Jaw, Da-Wei and Huang, Shih-Chia and Hwang, Jenq-Neng},
  journal=TIP,
  volume={27},
  number={6},
  pages={3064--3073},
  year={2018},
  publisher={IEEE}
}

@article{SnowFormer,
  title={SnowFormer: Scale-aware Transformer via Context Interaction for Single Image Desnowing},
  author={Chen, Sixiang and Ye, Tian and Liu, Yun and Chen, Erkang and Shi, Jun and Zhou, Jingchun},
  journal={arXiv preprint arXiv:2208.09703},
  year={2022}
}

@InProceedings{Attentive-GAN,
    author = {Qian, Rui and Tan, Robby T. and Yang, Wenhan and Su, Jiajun and Liu, Jiaying},
    title = {Attentive Generative Adversarial Network for Raindrop Removal From a Single Image},
    booktitle = CVPR,
    month = {June},
    year = {2018}
}

@article{NafNet,
  title={Simple Baselines for Image Restoration},
  author={Chen, Liangyu and Chu, Xiaojie and Zhang, Xiangyu and Sun, Jian},
  journal={arXiv preprint arXiv:2204.04676},
  year={2022}
}

@inproceedings{MPRNet,
    title={Multi-Stage Progressive Image Restoration},
    author={Syed Waqas Zamir and Aditya Arora and Salman Khan and Munawar Hayat
            and Fahad Shahbaz Khan and Ming-Hsuan Yang and Ling Shao},
    booktitle=CVPR,
    year={2021}
}

@InProceedings{HINet,
    author    = {Chen, Liangyu and Lu, Xin and Zhang, Jie and Chu, Xiaojie and Chen, Chengpeng},
    title     = {HINet: Half Instance Normalization Network for Image Restoration},
    booktitle = CVPR,
    month     = {June},
    year      = {2021},
    pages     = {182-192}
}

@InProceedings{Uformer,
    author    = {Wang, Zhendong and Cun, Xiaodong and Bao, Jianmin and Zhou, Wengang and Liu, Jianzhuang and Li, Houqiang},
    title     = {Uformer: A General U-Shaped Transformer for Image Restoration},
    booktitle = CVPR,
    month     = {June},
    year      = {2022},
    pages     = {17683-17693}
}

@article{ProRes,
      title={ProRes: Exploring Degradation-aware Visual Prompt for Universal Image Restoration}, 
      author={Jiaqi Ma and Tianheng Cheng and Guoli Wang and Xinggang Wang and Qian Zhang and Lefei Zhang},
      journal={arXiv preprint arXiv:2306.13653},
      year={2023}
}

@article{prompt_in_prompt,
  title={Prompt-In-Prompt Learning for Universal Image Restoration},
  author={Li, Zilong and Lei, Yiming and Ma, Chenglong and Zhang, Junping and Shan, Hongming},
  journal={arXiv preprint arXiv:2312.05038},
  year={2023}
}

@article{instrcut-ipt,
      title={Instruct-IPT: All-in-One Image Processing Transformer via Weight Modulation}, 
      author={Yuchuan Tian and Jianhong Han and Hanting Chen and Yuanyuan Xi and Ning Ding and Jie Hu and Chao Xu and Yunhe Wang},
      year={2024},
      journal={arXiv preprint arXiv:2407.00676},
}

@article{DA-CLIP,
  title={Controlling Vision-Language Models for Universal Image Restoration},
  author={Luo, Ziwei and Gustafsson, Fredrik K and Zhao, Zheng and Sj{\"o}lund, Jens and Sch{\"o}n, Thomas B},
  journal={arXiv preprint arXiv:2310.01018},
  year={2023}
}

@article{DiffBIR,
      title={DiffBIR: Towards Blind Image Restoration with Generative Diffusion Prior}, 
      author={Xinqi Lin and Jingwen He and Ziyan Chen and Zhaoyang Lyu and Bo Dai and Fanghua Yu and Wanli Ouyang and Yu Qiao and Chao Dong},
      year={2024},
      journal={arXiv preprint arXiv:2308.15070},
}

@article{autodir,
  title={Autodir: Automatic all-in-one image restoration with latent diffusion},
  author={Jiang, Yitong and Zhang, Zhaoyang and Xue, Tianfan and Gu, Jinwei},
  journal={arXiv preprint arXiv:2310.10123},
  year={2023}
}

@article{mperceiver,
      title={Multimodal Prompt Perceiver: Empower Adaptiveness, Generalizability and Fidelity for All-in-One Image Restoration}, 
      author={Yuang Ai and Huaibo Huang and Xiaoqiang Zhou and Jiexiang Wang and Ran He},
      year={2024},
      journal={arXiv preprint arXiv:2312.02918},
}

@article{StableSR,
      title={Exploiting Diffusion Prior for Real-World Image Super-Resolution}, 
      author={Jianyi Wang and Zongsheng Yue and Shangchen Zhou and Kelvin C. K. Chan and Chen Change Loy},
      year={2024},
      journal={arXiv preprint arXiv:2305.07015},
}

@article{DiffRestorer,
      title={Diff-Restorer: Unleashing Visual Prompts for Diffusion-based Universal Image Restoration}, 
      author={Yuhong Zhang and Hengsheng Zhang and Xinning Chai and Zhengxue Cheng and Rong Xie and Li Song and Wenjun Zhang},
      year={2024},
      journal={arXiv preprint arXiv:2407.03636},
}

@article{chen2025janus,
  title={Janus-Pro: Unified Multimodal Understanding and Generation with Data and Model Scaling},
  author={Chen, Xiaokang and Wu, Zhiyu and Liu, Xingchao and others},
  journal={arXiv preprint arXiv:2501.17811},
  year={2025}
}

@article{ma2024janusflow,
      title={JanusFlow: Harmonizing Autoregression and Rectified Flow for Unified Multimodal Understanding and Generation}, 
      author={Yiyang Ma and Xingchao Liu and Xiaokang Chen and others},
      journal={arXiv preprint arXiv:2411.07975},
      year={2024}
}

@article{xiao2024omnigen,
  title={Omnigen: Unified image generation},
  author={Xiao, Shitao and Wang, Yueze and Zhou, Junjie and Yuan, Huaying and Xing, Xingrun and Yan, Ruiran and Wang, Shuting and Huang, Tiejun and Liu, Zheng},
  journal={arXiv preprint arXiv:2409.11340},
  year={2024}
}

@article{wu2025qwenimagetechnicalreport,
      title={Qwen-Image Technical Report}, 
      author={Chenfei Wu and Jiahao Li and Jingren Zhou and others},
      year={2025},
      journal={arXiv preprint arXiv:2508.02324},
}

@article{brooks2022instructpix2pix,
  title={InstructPix2Pix: Learning to Follow Image Editing Instructions},
  author={Brooks, Tim and Holynski, Aleksander and Efros, Alexei A},
  journal={arXiv preprint arXiv:2211.09800},
  year={2022}
}

@article{zhou2022coop,
    title={Learning to Prompt for Vision-Language Models},
    author={Zhou, Kaiyang and Yang, Jingkang and Loy, Chen Change and Liu, Ziwei},
    journal=IJCV,
    year={2022}
}

@article{lorair,
      title={LoRA-IR: Taming Low-Rank Experts for Efficient All-in-One Image Restoration},
      author={Ai, Yuang and Huang, Huaibo and He, Ran},
      journal={arXiv preprint arXiv:2410.15385},
      year={2024}
}

@article{waveft,
      title={Exploring Sparsity for Parameter Efficient Fine Tuning Using Wavelets}, 
      author={Ahmet Bilican and M. Akın Yılmaz and A. Murat Tekalp and R. Gökberk Cinbiş},
      year={2025},
      journal={arXiv preprint arXiv:2505.12532},
}

@InProceedings{clip,
  title     = {Learning Transferable Visual Models from Natural Language Supervision},
  author    = {Radford, Alec and Kim, Jong Wook and Hallacy, Chris and Ramesh, Aditya and Goh, Gabriel and Agarwal, Sandhini and Sastry, Girish and Askell, Amanda and Mishkin, Pamela and Clark, Jack and others},
  booktitle = ICML,
  pages     = {8748--8763},
  year      = {2021},
}

@Article{T5,
  title   = {Exploring the Limits of Transfer Learning with a Unified Text-to-Text Transformer},
  author  = {Raffel, Colin and Shazeer, Noam and Roberts, Adam and Lee, Katherine and Narang, Sharan and Matena, Michael and Zhou, Yanqi and Li, Wei and Liu, Peter J.},
  journal = JMLR,
  volume  = {21},
  number  = {140},
  pages   = {1--67},
  year    = {2020},
}

@article{step1xedit,
  title={Step1X-Edit: A Practical Framework for General Image Editing},
  author={Liu, Shiyu and Han, Yucheng and Xing, Peng and others},
  journal={arXiv preprint arXiv:2504.17761},
  year={2025}
}

@misc{flux2,
    author={Black Forest Labs},
    title={{FLUX.2: Frontier Visual Intelligence}},
    year={2025},
    howpublished={\url{https://bfl.ai/blog/flux-2}},
}

@article{longcat,
  title={LongCat-Image Technical Report},
  author={{Meituan LongCat Team}},
  journal={arXiv preprint arXiv:2512.07584},
  year={2025}
}

@article{realrestorer,
  title={RealRestorer: Towards Generalizable Real-World Image Restoration with Large-Scale Image Editing Models},
  author={Yang, Yufeng and Zeng, Xianfang and Jiang, Zhangqi and others},
  journal={arXiv preprint arXiv:2603.25502},
  year={2026}
}

@inproceedings{lol,
  title={Deep Retinex Decomposition for Low-Light Enhancement},
  author={Wei, Chen and Wang, Wenjing and Yang, Wenhan and Liu, Jiaying},
  booktitle=BMVC,
  year={2018}
}

@inproceedings{lpips,
  title={The Unreasonable Effectiveness of Deep Features as a Perceptual Metric},
  author={Zhang, Richard and Isola, Phillip and Efros, Alexei A. and Shechtman, Eli and Wang, Oliver},
  booktitle=CVPR,
  year={2018}
}

@INPROCEEDINGS{div2k,
  author={Agustsson, Eirikur and Timofte, Radu},
  booktitle=CVPRW, 
  title={NTIRE 2017 Challenge on Single Image Super-Resolution: Dataset and Study}, 
  year={2017},
  volume={},
  number={},
  pages={1122-1131},
  }

@inproceedings{ARCNN,
author = {Dong, Chao and Deng, Yubin and Loy, Chen Change and Tang, Xiaoou},
title = {Compression Artifacts Reduction by a Deep Convolutional Network},
year = {2015},
isbn = {9781467383912},
publisher = {IEEE Computer Society},
address = {USA},
booktitle = ICCV,
pages = {576–584},
numpages = {9},
}

@INPROCEEDINGS{rain100l,
  author={Yang, Wenhan and Tan, Robby T. and Feng, Jiashi and Liu, Jiaying and Guo, Zongming and Yan, Shuicheng},
  booktitle=CVPR, 
  title={Deep Joint Rain Detection and Removal from a Single Image}, 
  year={2017},
  pages={1685-1694}}

@article{tekalp2026imagemodeling,
  author  = {A. Murat Tekalp},
  title   = {Image Modeling: From Handcrafted Priors to Foundation Image Models},
  journal = {IEEE Signal Processing Magazine},
  year    = {2026},
  note    = {Accepted for publication, Aug. 2026}
}
\end{document}